\documentclass[12pt,epsf]{article}
\usepackage{graphicx}
\usepackage{amssymb}
\usepackage{amsmath}
\begin{document}

\def\a{\alpha}
\def\b{\beta}
\def\c{\varepsilon}
\def\d{\delta}
\def\e{\epsilon}
\def\f{\phi}
\def\g{\gamma}
\def\h{\theta}
\def\k{\kappa}
\def\l{\lambda}
\def\m{\mu}
\def\n{\nu}
\def\p{\psi}
\def\q{\partial}
\def\r{\rho}
\def\s{\sigma}
\def\t{\tau}
\def\u{\upsilon}
\def\v{\varphi}
\def\w{\omega}
\def\x{\xi}
\def\y{\eta}
\def\z{\zeta}
\def\D{\Delta}
\def\G{\Gamma}
\def\H{\Theta}
\def\L{\Lambda}
\def\F{\Phi}
\def\P{\Psi}
\def\S{\Sigma}

\def\o{\over}
\newcommand{\gsim}{ \mathop{}_{\textstyle \sim}^{\textstyle >} }
\newcommand{\lsim}{ \mathop{}_{\textstyle \sim}^{\textstyle <} }
\newcommand{\vev}[1]{ \left\langle {#1} \right\rangle }
\newcommand{\bra}[1]{ \langle {#1} | }
\newcommand{\ket}[1]{ | {#1} \rangle }
\newcommand{\EV}{ {\rm eV} }
\newcommand{\KEV}{ {\rm keV} }
\newcommand{\MEV}{ {\rm MeV} }
\newcommand{\GEV}{ {\rm GeV} }
\newcommand{\TEV}{ {\rm TeV} }
\def\diag{\mathop{\rm diag}\nolimits}
\def\Spin{\mathop{\rm Spin}}
\def\SO{\mathop{\rm SO}}
\def\O{\mathop{\rm O}}
\def\SU{\mathop{\rm SU}}
\def\U{\mathop{\rm U}}
\def\Sp{\mathop{\rm Sp}}
\def\SL{\mathop{\rm SL}}
\def\tr{\mathop{\rm tr}}

\newcommand{\beq}{\begin{equation}}   
\newcommand{\eeq}{\end{equation}}
\newcommand{\bea}{\begin{eqnarray}}   
\newcommand{\eea}{\end{eqnarray}}
\newcommand{\bear}{\begin{array}}  
\newcommand {\eear}{\end{array}}
\newcommand{\bef}{\begin{figure}}  
\newcommand {\eef}{\end{figure}}
\newcommand{\bec}{\begin{center}}  
\newcommand {\eec}{\end{center}}
\newcommand{\non}{\nonumber}  
\newcommand {\eqn}[1]{\beq {#1}\eeq}
\newcommand{\la}{\left\langle}  
\newcommand{\ra}{\right\rangle}
\newcommand{\ds}{\displaystyle}
\def\SEC#1{Sec.~\ref{#1}}
\def\FIG#1{Fig.~\ref{#1}}
\def\EQ#1{Eq.~(\ref{#1})}
\def\EQS#1{Eqs.~(\ref{#1})}
\def\GEV#1{10^{#1}{\rm\,GeV}}
\def\MEV#1{10^{#1}{\rm\,MeV}}
\def\KEV#1{10^{#1}{\rm\,keV}}
\def\lrf#1#2{ \left(\frac{#1}{#2}\right)}
\def\lrfp#1#2#3{ \left(\frac{#1}{#2} \right)^{#3}}


\baselineskip 0.7cm

\begin{titlepage}

\begin{flushright}
IPMU-11-0001  \\
\end{flushright}

\vskip 1.35cm
\begin{center}
{\large \bf
Why have supersymmetric particles not been observed?
}
\vskip 1.2cm
Fuminobu Takahashi$^a$
and 
Tsutomu T. Yanagida$^{a,b}$

\vskip 0.4cm

{\it $^a$Institute for the Physics and Mathematics of the Universe,
University of Tokyo, Kashiwa 277-8568, Japan}\\
{\it $^b$Department of Physics, University of Tokyo, Tokyo 113-0033, Japan}

\vskip 1.5cm

\abstract{ If low-energy supersymmetry is the solution to the
  hierarchy problem, it is a puzzle why supersymmetric particles have
  not been observed experimentally to date. We show that
  supersymmetric particles in the TeV region can be explained if the
  fundamental cut-off scale of the theory is smaller than the
  $4$-dimensional Planck scale and if thermal leptogenesis is the
  source of the observed baryon asymmetry.  The supersymmetric
  particles such as sfermions and gauginos are predicted to be in the
  TeV region, while the gravitino is the LSP with mass of
  $O(100)$\,GeV and is a good candidate for dark matter.  Interestingly, the cosmological
  moduli problem can be solved in the theory with the low cut-off scale.  }
\end{center}
\end{titlepage}

\setcounter{page}{2}

Supersymmetry (SUSY) has been known as the most plausible candidate
for the theory beyond the standard model (SM). If low-energy SUSY is
indeed the solution to the hierarchy problem, then the masses of SUSY
particles such as squarks, sleptons and gauginos are naively of
$O(100)$\,GeV or so, and we would expect to already be seeing evidence
of these particles.  However, the SUSY particles have not been
observed experimentally to date, which pushes the SUSY scale beyond
the expected value.  In fact, in the minimal supersymmetric standard
model (MSSM), large soft SUSY breaking masses of $O(1)$ TeV are
typically required in order to avoid conflict with the LEP bound on
the light Higgs boson mass. This already implies that the solution to
the hierarchy problem is not the major reason for the low-energy SUSY.
If the characteristic SUSY scale is indeed in the TeV region, there
must be another reason for the presence of the low-energy SUSY at the
TeV scale, since otherwise the SUSY is likely broken at a higher
scale in the landscape~\cite{Dine:2005yq}.

In this letter we argue that the TeV-scale SUSY can be understood in a
theory with a cut-off scale, $\Lambda$, one order of magnitude lower
than the Planck scale $M_p$, if thermal
leptogenesis~\cite{Fukugita:1986hr} is the source of the observed
baryon asymmetry.  As noted in Ref.~\cite{Takahashi:2010uw}, the
cosmological moduli problem~\cite{Coughlan:1983ci,Goncharov:1984qm}
can be beautifully solved in this framework, using the solution
proposed long ago by Linde~\cite{Linde:1996cx}.  Our theoretical
framework has interesting implications for collider experiments, dark
matter search experiments, and inflation models, which we shall
describe below.

Let us consider the SUSY mass spectrum. We assume gravity-mediated SUSY breaking and  introduce a pseudomodulus $S$,
which has a non-vanishing $F$-term, 
\beq
|F_S| \;=\;  \sqrt{3}\,m_{3/2} M_p,
\eeq
where we have required the vanishingly small cosmological constant.
Given that the fundamental cut-off scale of the theory is  $\Lambda$, any non-renormalizable operators should
be suppressed by some powers of $\Lambda$.
Then the scalars acquire a mass from 
\beq
{\cal L} \;=\; - \int d^4 \theta\,  \frac{S^\dag S Q^\dag Q}{\Lambda^2},
\label{sfermion}
\eeq
where $Q$ collectively denotes the matter fields in the visible MSSM sector. The MSSM gauginos acquire a mass from
\beq
{\cal L} \;=\; - \int d^2 \theta\, \frac{S}{\Lambda} W_\alpha W_\alpha,
\label{gaugino}
\eeq
where $W_\alpha$ is a chiral superfield for the MSSM gauge multiplets. The 
scalar and gaugino masses are therefore given by
\beq
m_0\; \sim\; m_{1/2} \;\sim\; m_{3/2}\frac{M_p}{\Lambda}.
\label{susymass}
\eeq
The gravitino is generally lighter than the sfermion and the gauginos,
if the cut-off scale $\Lambda$ is lower than the Planck scale. As we
shall see below, $\Lambda$ must be one order of magnitude smaller than
the Planck scale to solve the cosmological moduli problem.  Thus, the
gravitino is the lightest SUSY particle (LSP), and is a good candidate
for dark matter. The little hierarchy between the soft SUSY breaking
masses and the gravitino mass is one of the important results in this
letter.

The gravitinos are produced by particle scatterings in thermal plasma,
and its abundance depends on the reheating temperature of the
Universe~\cite{Moroi:1993mb,Bolz:1998ek,Bolz:2000fu,Pradler:2006qh}
\begin{eqnarray}
    \label{eq:Yx-new}
    Y_{3/2} \;\simeq\; 1 \times 10^{-13} \left(1+\frac{m_{\tilde{g}}^2}{3 m_{3/2}^2} \right)
    \left( \frac{T_{R}}{10^{9}\ {\rm GeV}} \right),
   \end{eqnarray}
   where $m_{\tilde{g}}$ is the gluino mass evaluated at the
   reheating, $T_R$ denotes the reheating temperature, and we
   considered only the SU(3) contribution to the gravitino
   production. Using (\ref{susymass}), the gravitino density parameter
   is expressed by
\beq
\Omega_{3/2} h^2 \;\simeq\; 0.1\, c_3^2 \lrf{m_{3/2}}{100{\rm \,GeV}} \lrfp{\Lambda}{0.1 M_p}{-2} \lrf{T_R}{\GEV{9}},
\label{o32}
\eeq
where we have defined the gluino mass as $m_{\tilde{g}} = c_3 m_{3/2}
M_p/\Lambda$ with $c_3 = O(1)$, and we dropped the contribution of the
transverse component of the gravitino. Let us presume that thermal
leptogenesis is the source of the observed baryon asymmetry. Then
successful thermal leptogenesis requires $T_R \gtrsim
\GEV{9}$~\cite{Buchmuller:2004nz,Buchmuller:2005eh}.  Combined with
(\ref{o32}), the upper bound of the gravitino mass is then fixed to be
about $100$\,GeV in order to account for the dark matter abundance,
$\Omega_{DM} h^2 = 0.1123 \pm 0.0035$~\cite{Komatsu:2010fb}. The
SUSY-breaking mass scale can be pushed into the TeV region in order to
account for the observed baryon asymmetry and dark matter abundance
(See Eq.~(\ref{susymass})).  This explains why the SUSY particles have
not been observed experimentally to date, if the SUSY is
preferentially broken at a high scale~\cite{Dine:2005yq}. (See note added
for another argument.) 
We emphasize here that the requirement of thermal leptogenesis plays an
essential role in the above argument.

Next let us briefly show how the moduli problem can be solved; see
Ref.~\cite{Takahashi:2010uw} and references therein for details.  If
the theory has a fundamental cut-off scale $\Lambda$, there is
generically the following quartic coupling,
\beq
{\cal L} \;=\; - \int d^4 \theta\,  \frac{\chi^\dag \chi Z^\dag Z}{\Lambda^2},
\label{quartic}
\eeq
where $Z$ represents the modulus (including $S$), and $\chi$ denotes a
chiral superfield which dominates the energy density of the Universe
when $Z$ starts to oscillate. In the standard scenario, the $\chi$ is identified
with the inflaton. 
 If $\Lambda \sim 0.1\,M_p$, the
modulus has a mass of $O(10)H$, where $H$ is the Hubble parameter,
then the modulus $Z$ follows the time-dependent minimum and amplitude
of coherent oscillations is exponentially suppressed~\cite{Linde:1996cx}. 
Thus, the cosmological moduli problem is solved. 
This solution requires $\Lambda$ to be smaller than or equal to
 $0.1 M_p$. In order not to affect the successful grand unification, we consider
$\Lambda \sim 0.1 M_p$ in this letter.\footnote{The operator
  $\frac{<\Sigma>}{\Lambda} W_{\alpha}W_{\alpha}$ violates the GUT
  unification of gauge coupling constants if $\Lambda < 0.1M_p$. Here,
  the $\Sigma$ is the adjoint ${\bf 24}$ representation of $SU(5)_{\rm
    GUT}$.}

There is an important constraint on the reheating temperature for the above
mechanism to work.
The large Hubble-induced mass term disappears after the reheating, and so,
the decay rate of the $\chi$, $\Gamma_{\chi}$, should satisfy
$\Gamma_{\chi} \ll O(0.1) m_Z$, where $m_Z \sim O(10) m_{3/2}$ is the modulus mass.
This inequality is satisfied for the reference values, $m_{3/2} \sim 100$\,GeV
and $T_R \sim 10^9$\,GeV.
When the Hubble-induced mass term disappears at the reheating, the potential minimum 
is expected to change accordingly.
One may think that the modulus oscillations are then induced afterwards. However,
the modulus continues to follow the minimum during and after the reheating 
since its mass scale is larger than the Hubble parameter at that time, as long as the above inequality is satisfied.
We have numerically checked that the modulus amplitude is indeed suppressed enough
to solve the moduli problem, taking account of the effect of reheating.

It has been known that, for the gravitino LSP of mass $m_{3/2} \sim
100{\rm \,GeV}$, the next-to-lightest SUSY particle (NLSP) is
long-lived and decays into the SM particles and the gravitino during
big bang nucleosynthesis (BBN), which alters the light element
abundances in various ways~\cite{Pospelov:2006sc,
  Kawasaki:2004qu}. The BBN constraints on the NLSP can be avoided if
the R-parity is not an exact symmetry, but explicitly broken by a
small amount~\cite{Takayama:2000uz,Buchmuller:2007ui}. Such R-parity
violation may be ubiquitous in the string
landscape~\cite{Kuriyama:2008pv}.  In order to not erase the baryon
asymmetry, the size of the R-parity violation is
constrained~\cite{Campbell:1991at,Endo:2009cv}.  There is a certain
range of parameters where the NLSP decays well before the BBN while
the baryon asymmetry is not erased.  Note that the gravitino is
long-lived because of the Planck suppressed interactions even if the R-parity is broken,
and therefore becomes dark matter .  The gravitino decay may leave some signature
in the cosmic-ray
spectrum~\cite{Takayama:2000uz,Buchmuller:2007ui,Ibarra:2007wg,Ishiwata:2008cu},
which may be discovered in the future indirect dark matter search.

Lastly let us consider an implication for inflation models. With the
cut-off scale of the theory below the Planck scale, the inflaton mass
easily exceeds the Hubble parameter during inflation, because of the
following operator,
\beq
{\cal L}\;=\;-\int d^4 \theta \frac{|\phi|^4}{\Lambda^2},
\label{inf4}
\eeq
where $\phi$ denotes the inflaton.  Namely, the $\eta$-problem gets
worse than usual~\cite{Takahashi:2010uw}.  This problem can be
circumvented if the inflaton mass is protected by symmetry, such as
the shift symmetry.  Indeed there are such models that the inflaton
mass is forbidden by symmetry~\cite{Kawasaki:2000yn, Watari:2000jh,
  Takahashi:2010ky, Kallosh:2010ug}.

We have assumed that
the gravitinos are mainly produced by thermal scatterings. On the
other hand, the gravitinos are known to be non-thermally produced by
the inflaton decay~\cite{Kawasaki:2006gs,Endo:2007ih}, and such
non-thermal gravitino production should be suppressed.  This places
upper bounds on the inflaton mass and the vacuum expectation value
(VEV).  One successful inflation model is a chaotic inflation model
with a discrete symmetry. In this model, the inflaton mass is
protected by a shift symmetry, and the inflaton and its companion
field have vanishing VEVs because of the $Z_2$
symmetry~\cite{Kawasaki:2000yn}.  Therefore the model avoid the
$\eta$-problem and the non-thermal gravitino production does not
occur.

So far we have not specified the origin of the low cut-off scale. If
all the matter fields including the MSSM sector is confined on the
three-dimensional brane while the extra dimensions are compactified
with a typical radius larger than the higher-dimensional Planck length
$M_*$, the four-dimensional low-energy effective theory has a cut-off
scale $M_*$, which is lower than the four-dimensional Planck scale.
We identify the cut-off scale $\Lambda$ with the higher-dimensional
Planck length $M_*$.  Alternatively, there might be strong dynamics
near the Planck scale, which results in a large coupling rather than
the low cut-off scale, as proposed in Ref.~\cite{Takahashi:2010uw}.

To summarize, we have proposed that the fundamental cut-off scale
$\Lambda$ of the theory is lower than the Planck scale $M_p$, and have
shown that this gives an explanation for why the SUSY particles have
escaped the detection so far.  The typical SUSY scale can be pushed
into the TeV region, which ameliorates the constraints of the
flavor-changing and CP violation processes. The gravitino is the LSP
of mass $m_{3/2} \sim 100$\,GeV, and accounts for the observed dark
matter abundance. The requirement of successful thermal leptogenesis
plays an important role to reach the above conclusion.

There are interesting implications. The BBN constraints on the NLSP
can be avoided if the R-parity is explicitly broken. Then the
gravitino dark matter is unstable and decays into the SM particles,
which may leave observable signature in the cosmic-ray
spectrum~\cite{Buchmuller:2007ui,Ibarra:2007wg,Ishiwata:2008cu}. The
R-parity violation may be also seen at LHC~\cite{Aad:2009wy}. The inflation
model should be such that the inflaton mass is protected by
symmetry. One example is the chaotic inflation with a discrete
symmetry.  Because of the rich implications, our proposal can be
tested by collider, dark matter experiments as well as the CMB
observation in the near future.

\vspace{5mm}
 {\it Note added:}
As mentioned in the text, we have found that the following inequality must be met for
the Linde's solution to the moduli problem to work:
\beq
m_{3/2} \; \gtrsim 100{\rm\,GeV}  \lrfp{T_R}{2\times \GEV{9}}{2}.
\eeq
Thus, if we require that the leptogenesis is the source of the baryon asymmetry, 
$T_R$ must be higher than $2\times 10^9$\,GeV~\cite{Buchmuller:2004nz}, which then leads to
a lower bound on the gravitino mass, $m_{3/2} \gtrsim 100$\,GeV.
Therefore the soft SUSY breaking masses are pushed into the TeV region or heavier (see Eq.~(\ref{susymass})), 
which explains why the
SUSY particles have not been discovered to date. If the above inequality is not satisfied,
the modulus would dominate the energy density of the Universe and produces
huge entropy at the decay, which dilutes the pre-existing baryon asymmetry.
Since the suppression of the modulus amplitude is exponentially sensitive to the
above condition, this argument provides us with a sharp lower bound on the
SUSY breaking scale.

\section*{Acknowledgment}

This work was supported by the Grant-in-Aid for Scientific Research on 
Innovative Areas (No. 21111006) [FT],  Scientific Research (A)
(No. 22244030 [FT] and 22244021 [TTY]), and JSPS Grant-in-Aid for Young Scientists (B) (No. 21740160) [FT]. 
This work was also supported by World Premier
International Center Initiative (WPI Program), MEXT, Japan.


\end{document}